\documentclass[12pt]{article}

\usepackage{axodraw,epsfig,cite}

\input paperdef

\evensidemargin \oddsidemargin
\begin{document}

%%%%%%%%%%%%%%%%%%%%%%%%%%%%%%%%%%%%%%%%%%%%%%%%%%%%%%%%%%%%%%
%%%%%%%%%%%%%%%%%%%%%%%%%%%%%%%%%%%%%%%%%%%%%%%%%%%%%%%%%%%%%%

\thispagestyle{empty}
\setcounter{page}{0}
\def\thefootnote{\fnsymbol{footnote}}

\mbox{}\vspace{-1em}
\begin{flushright}
BNL--HET--02/9\\
hep-ph/0203067 \\
 \end{flushright}

\vspace{2em}

\begin{center}

{\large\sc {\bf The Higgs Boson Production Cross Section}}

\vspace*{0.4cm} 

{\large\sc {\bf as a Precision Observable?}}

\vspace{1cm}

{\sc S.~Dawson$^{\,1}$%
\footnote{
email: dawson@bnl.gov
}%
~and S.~Heinemeyer$^{\,1,2}$%
\footnote{
email: Sven.Heinemeyer@physik.uni-muenchen.de
}%
}

\vspace*{1cm}

$^1$ Physics Dept., Brookhaven Natl.\ Lab., Upton, NY 
11973, USA

\vspace*{0.4cm}

$^2$Institut f\"ur theoretische Elementarteilchenphysik, LMU M\"unchen,\\ 
Theresienstr.\ 37, D-80333 M\"unchen, Germany

\end{center}

\vspace*{1cm}

\begin{abstract}

We investigate what can be learned at a linear collider
about the sector of electroweak
symmetry breaking from a precise measurement of the Higgs boson
production cross section  through
the process \eetohZ .
 We focus on deviations from the Standard
Model arising in its minimal supersymmetric extension. The analysis is
performed within two
realistic future scenarios, taking into account all prospective
experimental errors on supersymmetric particle masses as well as
uncertainties from unknown higher order corrections. We find that
information on $\tb$ and $\MA$ could be obtained from a cross section
measurement with a precision
of $0.5 - 1\%$. Alternatively, information could be obtained
on the gaugino mass parameters $M_2$ and $\mu$ if they are relatively
small, $M_2, \mu \approx 200 \gev$.

\end{abstract}

\def\thefootnote{\arabic{footnote}}
\setcounter{footnote}{0}

\newpage

%%%%%%%%%%%%%%%%%%%%%%%%%%%%%%%%%%%%%%%%%%%%%%%%%%%%%%%%%%%%%%
%%%%%%%%%%%%%%%%%%%%%%%%%%%%%%%%%%%%%%%%%%%%%%%%%%%%%%%%%%%%%%

\section{Introduction}

One of the fundamental problems facing
particle physics is understanding the nature of electroweak
symmetry breaking. If this symmetry breaking is due
to a light Higgs boson, then the Higgs boson
will certainly be discovered at the Tevatron~\cite{higgstev} 
or the LHC~\cite{atlastdr,cms,ehow}. The remaining challenge 
will then be to understand whether this new  object is the 
Higgs boson of the Standard Model (SM) or some more exotic
particle.  In the SM, the couplings of the Higgs boson to
all particles are completely fixed once the mass is known and
so the validity of the SM can be confirmed by measuring
Higgs production and decay rates and eventually the Higgs potential
itself~\cite{hhh}. In alternative models,
the Higgs couplings can be quite different from the SM values and so can
potentially be used to distinguish between models.

A linear collider (LC) with an
energy in the range $\sqrt{s} \sim 350-500 \gev$ has the capability 
for performing precision measurements of both Higgs boson production
and decay rates~\cite{teslatdr,orangebook,acfarep}, provided that the
Higgs boson mass, $\Mh$, lies in the range predicted by electroweak
precision observables, $\Mh \lsim 200 \gev$~\cite{lepewwg,blueband}.
The dominant production mechanism for such a light Higgs boson is 
\eetohZ~\cite{hprod}, with the largest decay channel being
 $\hbb$ or $h \to WW^*$. The measurements must be 
interpreted in terms of SM expectations or some model of physics 
beyond the SM. The goal is then to use the experimental data
to disentangle the underlying structure of the model.  
An important question is thus the required experimental precision 
for production rates and branching ratios in order to distinguish it
from the SM and perhaps to measure the parameters of the new theory.

An integrated luminosity of $\cL \sim 500$~\ifb\ and 
$\sqrt{s} = 350 \gev$ is expected to produce measurements of the
various Higgs branching ratios with precisions 
in the $2-10\%$ range at an $e^+e$
collider~\cite{teslatdr,orangebook,acfarep}. 
The precision will be less at $\sqrt{s}=500 \gev$, primarily
due to the reduced rate~\cite{teslatdr,brau}.
The LHC can measure some, but not all, Higgs branching ratios, with a 
precision which is typically less than that obtainable at a 
LC~\cite{zeppi}.

Supersymmetric (SUSY) theories~\cite{mssm} are widely considered as the
theoretically most appealing extension of the SM. They are
consistent with the approximate unification of the three gauge coupling
constants at the GUT scale and provide a way to cancel the quadratic
divergences in the Higgs sector, hence stabilizing the huge hierarchy between
the GUT and the Fermi scales. Furthermore, in SUSY theories the breaking
of the electroweak symmetry is naturally induced at the Fermi scale,
and the lightest supersymmetric particle can be neutral, weakly interacting
and absolutely stable, providing  a natural solution for the
Dark Matter problem.
Therefore the
implications of the measurements of Higgs boson branching 
ratios have been extensively studied within the context of
the 
minimal supersymmetric model
(MSSM)~\cite{higgsbrimp,higgsbrimp2}.
This model is extremely predictive and so is useful for
comparing the experimental reach achievable in various channels.

In this note we address the question of whether the total \eetohZ\
cross section, $\sihZ$, can be used as a precision observable
to help determine the structure of the electroweak sector of the
MSSM. 
 The measurement of the \eetohZ\ Higgsstrahlung
production cross 
section  is expected to achieve a $2-3\%$ accuracy at 
$\sqrt{s} = 350 \gev$~\cite{tesl atdr}. This assumes $\cL = 500$~\ifb\
and the analysis of the $Z \to l^+l^-$ events only.
From this measurement, some restrictions can be inferred about the
parameters of the MSSM, which we investigate here. For the
Higgsstrahlung process, the complete next-to-leading order corrections
(involving SM and SUSY particles), including all vertex and box
corrections have been calculated~\cite{eehZ1l}. More recently also the
leading \twol\ corrections have been included~\cite{eehZ2l}.
Since these corrections are significant, their inclusion is 
crucial for drawing conclusions about the underlying model.

Our study differs considerably from previous studies of the Higgs
branching ratios in that we  investigate plausible future
scenarios and estimate uncertainties from all relevant sources.
 We assume that some MSSM particle masses and mixing
angles have been determined at the LHC and/or the LC, and
vary {\em all} inputs accordingly within realistic errors, instead of
fixing all parameters and then varying just one or two.
Furthermore, the anticipated theory errors from unknown higher order
corrections in the MSSM Higgs sector are taken into account in a
consistent manner. 
We then ask what can
be learned about the remaining unknown parameters of the model.
These assumptions try to represent {\em possible} future scenarios and
thus give an idea of what might be inferred from a precise $\sihZ$
measurement. 

The rest of this paper is organized as follows: In section~2 we review
the necessary MSSM input parameters and existing 
higher order corrections in the
Higgs sector. Our approach to  the investigation, with emphasis on the
attempt to look into realistic future scenarios, is explained in
detail in section~3. Section~4 contains our analysis and the
corresponding results, while  conclusions can be found in section~5.

%%%%%%%%%%%%%%%%%%%%%%%%%%%%%%%%%%%%%%%%%%%%%%%%%%%%%%%%%%%%%%
%%%%%%%%%%%%%%%%%%%%%%%%%%%%%%%%%%%%%%%%%%%%%%%%%%%%%%%%%%%%%%

\section{The MSSM: Basics}

The Higgs sector of the MSSM consists of two Higgs doublets,
$\cHe$ and $\cHz$~\cite{hhg}.  After electroweak symmetry breaking, there
remain $5$ physical Higgs bosons:  $h, H, A$, and $H^\pm$. 
In this note, we will be concerned only with the production of the
lightest Higgs boson,~$h$.
The Higgs sector is described at tree level by two
additional parameters (besides the SM parameters),
 which are usually chosen to be
$\tb$, the ratio of the Higgs VEVs, and $\MA$, the mass of the
pseudoscalar Higgs boson. The mass eigenstates of the neutral scalar
Higgs bosons are obtained from  the interaction eigenstates
$\phi_1$ and $\phi_2$ by the rotation,
\BE
\VL H \\ h \VR = \ML \Ca & \Sa \\ -\Sa & \Ca \MR \; 
                 \VL \phi_1 \\ \phi_2 \VR ,
\EE
where at tree level
\BE
\tan 2\al = \tan 2\be \KL \frac{\MA^2+\MZ^2}{\MA^2-\MZ^2} \KR .
\EE
At tree level, the
 mass of the lightest Higgs boson is completely fixed in terms
of $\MZ$, $\MA$ and $\tb$.

\smallskip
The process \eetohZ\ proceeds (at the tree-level) via the Feynman diagram
shown in \reffi{eezhfd} and is hence sensitive to the $ZZh$ coupling.
At tree level, the $ZZh$ coupling in the MSSM is altered from the SM
value, 
\BE
g_{ZZh}^{\SU}= g_{ZZh}^{\SM} \; \Sba .
\EE
For $\MA \gg \MZ$, $\Sba \to 1$ and the coupling of the lightest MSSM
Higgs boson to the $Z$ boson approaches that of the SM.  We therefore
expect that  $\sihZ^{\SU}$ will be sensitive to small $\MA$.  

%%%%%%%%%%%% F I G U R E %%%%%%%%%%%%%%%%%%%%%%%%%%%%%%%%%%%%%
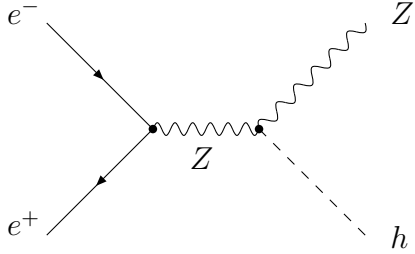
\begin{figure}[htb!]
\BC
\begin{picture}(100,100) 
\SetScale{0.8}
\ArrowLine(0,100)(50,50)
\ArrowLine(50,50)(0,0)
\Vertex(50,50){2}
\Photon(50,50)(100,50){3}{6}
\Vertex(100,50){2}
\Photon(100,50)(150,100){3}{6}
\DashLine(100,50)(150,0){5}
\put(54,25){$Z$}
\put(-15,80){$e^-$}
\put(-15,0){$e^+$}
\put(130,-5){$h$}
\put(130,80){$Z$}
\SetScale{1}
\end{picture}
\vspace{.5em}
\caption{Feynman diagram for lowest order contribution to
\eetohZ.}
\label{eezhfd}
\EC
\end{figure}
%%%%%%%%%%%% F I G U R E %%%%%%%%%%%%%%%%%%%%%%%%%%%%%%%%%%%%%

There are two important effects which arise when going beyond the tree
level. 
The first is that the Higgs boson mass prediction is significantly
increased by radiative corrections, leading to an upper bound at the
\twol\ level~\cite{mhiggs2lA,mhiggs2lB,bse} of $\Mh \lsim 135
\gev$~\cite{mhiggs2lB}. The most important
corrections are those in the $t/\Stop$~sector~\cite{mhiggs1l} and for
large $\tb$ also those in the $b/\Sbot$~sector.
The mass matrices in the basis of the current eigenstates $\StopL,
\StopR$ and $\SbotL, \SbotR$ are given by
\BEA
\label{stopmassmatrix}
{\cal M}^2_{\Stop} &=&
  \ML \MstL^2 + \mt^2 + \CZb (\edz - \frac{2}{3} \sw^2) \MZ^2 &
      \mt \Xt \\
      \mt \Xt &
      \MstR^2 + \mt^2 + \frac{2}{3} \CZb \sw^2 \MZ^2 
  \MR, \\
&& \non \\
\label{sbotmassmatrix}
{\cal M}^2_{\Sbot} &=&
  \ML \MsbL^2 + \mb^2 + \CZb (-\edz + \frac{1}{3} \sw^2) \MZ^2 &
      \mb \Xb \\
      \mb \Xb &
      \MsbR^2 + \mb^2 - \frac{1}{3} \CZb \sw^2 \MZ^2 
  \MR,
\EEA
where $\sw^2 = 1 - \cw^2 = 1 - \MW^2/\MZ^2$ and  
\BE
\mt \Xt = \mt (\At - \mu \CTb) , \quad
\mb\, \Xb = \mb\, (\Ab - \mu \Tb) .
\label{eq:mtlr}
\EE
Here $\At$ denotes the trilinear Higgs--stop coupling, $\Ab$ is the
Higgs--sbottom coupling, and $\mu$ is the Higgs mixing parameter.
SU(2) gauge invariance leads to the relation
\BE
\MstL = \MsbL .
\EE
The two mass matrices (\ref{stopmassmatrix}), (\ref{sbotmassmatrix})
are diagonalized by the angles $\tst$ and $\tsb$, 
respectively. The physical squark masses are  $\mste$, $\mstz$,
$\msbe$ and $\msbz$. 
Specifying $\mste$, $\mstz$, and $\tst$, along with $\mu$ and $\tb$
therefore
implicitly fixes the tri-linear mixing parameter $\At$, and similarly in
the $b$-squark sector.   
The radiatively corrected value for the lightest MSSM Higgs boson
mass, $\Mh$, depends sensitively on the
parameters of the stop mass matrix~(\ref{stopmassmatrix}).
   
The second important effect of going beyond tree level is that the
SUSY particles enter into loop corrections. The complete set of
one-loop corrections to the process \eetohZ\ has been computed in
\citere{eehZ1l}.   
In addition, the leading two loop corrections have been
included~\cite{eehZ2l}. The effects of including these corrections
have been discussed in detail in \citere{eehZ2l} and are seen to be
large. This applies in particular for the \twol\
corrections. Our analysis includes therefore all one-loop
SM and SUSY corrections, along with the leading two-loop
corrections. 
From the analysis in \citere{eehZ2l} one can infer a
theoretical uncertainty due to unknown higher order corrections for
the prediction of $\sihZ$ of $\sim 5\%$.

%%%%%%%%%%%%%%%%%%%%%%%%%%%%%%%%%%%%%%%%%%%%%%%%%%%%%%%%%%%%%%
%%%%%%%%%%%%%%%%%%%%%%%%%%%%%%%%%%%%%%%%%%%%%%%%%%%%%%%%%%%%%%

\section{Concept of the analysis}
\label{sec:concept}

The focus here is to determine in the context of SUSY what new
information can be obtained from a precision measurement of 
$\sihZ$, beyond the direct measurement of the  lightest Higgs
boson mass.
At the time of a $\sihZ$~measurement at the LC, SUSY (if it exists
at a low mass scale) will have been
discovered at the LHC and possibly confirmed by the LC. Therefore some
SUSY parameters will be known with high precision from the LC
measurements, while others (e.g. masses beyond the kinematic reach of the
LC) will be known with lesser precision from the LHC data. In the
Higgs sector it is possible that only the lightest MSSM Higgs boson
will have been measured (e.g.\ for $\MA \gsim 300$ and moderate $\tb$
values, $\tb \sim 10$)~\cite{atlastdr,cms,teslatdr}. Only for
relatively small masses, $\MH, \MA \lsim \sqrt{s}/2$, will the heavy Higgs
bosons be visible at the LC. 

In a realistic analysis at the time of the LC the following has to
be taken into account:
\begin{itemize}
\item uncertainties of the measured SM parameters
\item uncertainties of the measured MSSM parameters
\item intrinsic uncertainties on the theoretical prediction of the
      MSSM Higgs sector parameters ($\Mh, \sihZ, \ldots$) from unknown
higher order corrections
\item bremsstrahlung
\item beamstrahlung
\item other machine related uncertainties, e.g.\ due to the luminosity
      measurement, detector smearing etc.
\end{itemize}

A full simulation clearly goes beyond the scope of this exploratory
analysis. However, we try to give a realistic impression about the
information which can be obtained from a $\sihZ$ measurement. To this
end we include the following:
\begin{itemize}
\item the relevant SM uncertainties arising from the $\mt$ 
      measurement
\item we take into account {\em all} uncertainties on the MSSM parameters
      from their measurement at the LHC~\cite{atlastdr,cms} 
and/or the LC~\cite{teslatdr,orangebook,acfarep}%
\footnote{
The errors are similar to those used in \citere{gigaz}, where besides
the pure experimental resolution also the anticipated theoretical
uncertainty entering the extraction of the parameters has been taken
into account.
}%
.
To study the dependence of the cross section on the parameters, 
we vary {\em all} parameters within their expected precisions and 
include effects of SUSY particles beyond the leading order as 
described in the previous section.
\item we assume a future theoretical uncertainty in the prediction of 
      $\Mh$ from the other SUSY input parameters of 
$0.5 \gev$ (which affects mainly the connection of the different SUSY
parameters to each other). For the theoretical prediction of $\sihZ$
an uncertainty of~1\% is assumed from unknown higher order
corrections. However, the Higgs boson mass value that will be used in
the future will be determined to $\pm 0.05 \gev$ (see below) and thus
will have a negligible error.
(Numerically the uncertainty of $\sihZ$ is taken into account by
allowing a variation of the Higgs boson mass as an input parameter in
the $\sihZ$ evaluation by $\pm 0.5 \gev$. This (by numerical
coincidence) reproduces the ``desired'' theoretical uncertainty in
$\sihZ$ of $\sim 1\%$.)
\item we do not include beamstrahlung, bremsstrahlung, or detector 
      effects, which are beyond the scope of this note. While the
latter can only be realized in a full simulation, the former mostly
induce a shift in the numerical results, but have a much smaller
effect on the errors.
\item we neglect luminosity errors. Concerning these, it might be
      helpful not to investigate $\sihZ$ directly, but to consider
e.g.\ $\sihZ/\si_{ZZ}$, since in this ratio many
uncertainties cancel out. However, the
idea of this analysis is to show the possible potential of a precise
cross section measurement, which can already be obtained from an
analysis of $\sihZ$ alone.
\end{itemize}

Taking into account the relevant uncertainties in the above manner
necessarily weakens the potential of a precise $\sihZ$
measurement, see \refse{sec:analysis}. 
This approach is contrary to existing analyses~\cite{higgsbrimp2}. In
these previous analyses, all parameters, except for the
one under investigation, are fixed. Furthermore, all theoretical
uncertainties for the evaluation of the Higgs sector observables are
neglected. The potentially measured effect is then  attributed solely to
the one parameter under investigation, whereas part of the effect
could be due to  other sources, such as variations in one of
the parameters held fixed (within the
corresponding experimental errors) or due to the theoretical
uncertainties. In this way the sensitivity to the investigated
parameters is incorrectly enhanced. Our approach, on the other hand,
results in a smaller sensitivity, but constitutes a more 
realistic scenario for the investigation of LC analyses.

\bigskip
For this analysis we assume that $\sihZ$ is
measured at $\sqrt{s} = 350 \gev$ with $\cL = 500$~\ifb.%
\footnote{
Possible LC run scenarios have been investigated in
\citere{runscen}. They usually assume first some high(er)-energy run
and afterwards several shorter runs at lower energies, which we
summarize here as one run at $\sqrt{s} = 350 \gev$ with 
$\cL = 500$~\ifb. 
}%
~In all the investigated scenarios, we assume that the Higgs boson mass
will have been measured to an experimental accuracy 
of~\cite{teslatdr,orangebook,acfarep} 
\BE
\Mh^{\rm exp} = 115 \pm 0.05 \gev .
\EE
However, as mentioned above, within the MSSM this experimental error
will always be dominated by the theoretical uncertainty on
the prediction of $\Mh$ due to unknown
higher order corrections.  While the current
uncertainty in the $\Mh$ prediction is estimated to be 
$\sim 3 \gev$~\cite{dmhtheo}, we assume for the future uncertainty
\BE
\de\Mh^{\rm theo} ~({\rm future}) = \pm 0.5 \gev.
\EE
Also the dependence of $\Mh$ on the top quark mass is very strong,
$\de\mt/\de\Mh \approx~1$. However, $\mt$ will be determined
to an accuracy better
than $\sim 130 \mev$ at a LC~\cite{teslatdr,mtdet}, so that the
parametric uncertainty is smaller than the theoretical uncertainty. It
is, however, taken into account. 

Since the value of $\Mh$ in the MSSM is not a free parameter, but
depends on the other SUSY parameters, they have to be chosen such that
the value of $\Mh = 115 \pm 0.5 \gev$ emerges. The numerical
evaluation of the MSSM Higgs sector (including $\Mh$ and $\sihZ$) is
based on the code \fhxs~\cite{feynhiggs,eehZ2l}.

%%%%%%%%%%%%%%%%%%%%%%%%%%%%%%%%%%%%%%%%%%%%%%%%%%%%%%%%%%%%%%
%%%%%%%%%%%%%%%%%%%%%%%%%%%%%%%%%%%%%%%%%%%%%%%%%%%%%%%%%%%%%%

\section{Analysis and results}
\label{sec:analysis}

In order to make progress in understanding the sensitivity of the total
cross section to the input parameters, 
the approach explained above has been applied to two {\em possible}
future scenarios. In both scenarios we make assumptions what parameters
will be measured and what parameters are left free. This choice,
since it involves the unknown MSSM parameters and their detectability,
is of course subject to personal opinions. However, the scenarios
certainly reflect the possible strength of the $\sihZ$ measurement as
explained in the previous section. 

\subsection{The Higgs sector scenario}
\label{subsec:higgsscenario}

In the first scenario we assume that the gaugino and squark 
masses and mixing angles
have been measured at the LHC~\cite{atlastdr,cms} and/or the
LC~\cite{teslatdr,orangebook,acfarep}. For our analysis, the most
important input parameter is the top quark mass and its associated error.
Here we assume 
\BE
\mt^{\rm exp} = 175 \pm 0.1 \gev~,
\EE
which is the anticipated precision from a  high energy linear 
collider~\cite{mtdet}.

In the $\Stop$~sector, we chose
\BEA
\mste &=& 500 \pm 2 \gev \cr
\mstz &=& 700 \pm 10 \gev \cr
\sintt &=& -.69 \pm 0.014~.
\label{stopunc}
\EEA
This precision for $\mste$ and $\sintt$ could most probably only be
realized with an LC measurement at an energy of $\sqrt{s}=1 \tev$. 
A more conservative choice would be $\de\mste = 10 \gev$ and an error
on $\sintt$ of up to 10\%, which can be achieved at the
LHC~\cite{atlastdr}%
\footnote{
The precision of 10\% for $\sintt$ is relatively preliminary and
optimistic. The subject of $\Stop$~measurements at the LHC is still
under development, see e.g.~\citere{LHCstop}. 
}%
.
In the analysis we will first investigate the implications of a
LC precision, but comment also on the LHC precision results as well.

With the above measurements, $\At$ is given implicitly in terms of
$\mste$, $\mstz$, $\sintt$, $\mu$ and $\tb$.  We furthermore fix,
\BE
\mu = 200 \pm 1 \gev.
\EE
We assume approximate unification of the trilinear Higgs-sfermion
couplings and take: 
\BEA
\label{Ab}
\Ab &=& \At \pm 10\% \\
A_l &=& A_t \pm 1\% ~.
\EEA
In addition, we assume the relationship between gaugino masses predicted
in many unified models. The specific values we take are:
\BEA
M_2 &=& 400 \pm 2 \gev \cr
M_1 &=& {5\over 3} \, {\sw^2\over\cw^2} \, M_2 \pm 1 \gev \cr
\mgl = M_3 &=& 500 \pm 10 \gev ~.
\EEA
Finally, for the remaining sfermion sector we choose 
\BEA
\label{MSbotR}
\MsbR &=& \MstR \pm 10\% \\
\msee, \msez &=& 200 \pm 2 \gev ~,
\EEA
where the selectron masses enter in the vertex and box corrections.   
The uncertainties chosen above are consistent with those given
in \citeres{atlastdr,cms,teslatdr}, see \refse{sec:concept}.
The \refeqs{Ab} and (\ref{MSbotR}) reflect the assumed future
measurement of the scalar bottom sector. However, the $b/\Sbot$~sector
plays only a minor role here, since (as will be shown below) either
$\mu$ or $\tb$ (or both) do not reach large values. This, however, is
necessary to have large corrections from $b/\Sbot$~loops to the MSSM
Higgs sector.

%%%%%%%%%%%% F I G U R E %%%%%%%%%%%%%%%%%%%%%%%%%%%%%%%%%%%%%
\begin{figure}[htb!]
\begin{center}
\begin{tabular}{p{0.48\linewidth}p{0.48\linewidth}}
\mbox{\epsfig{file=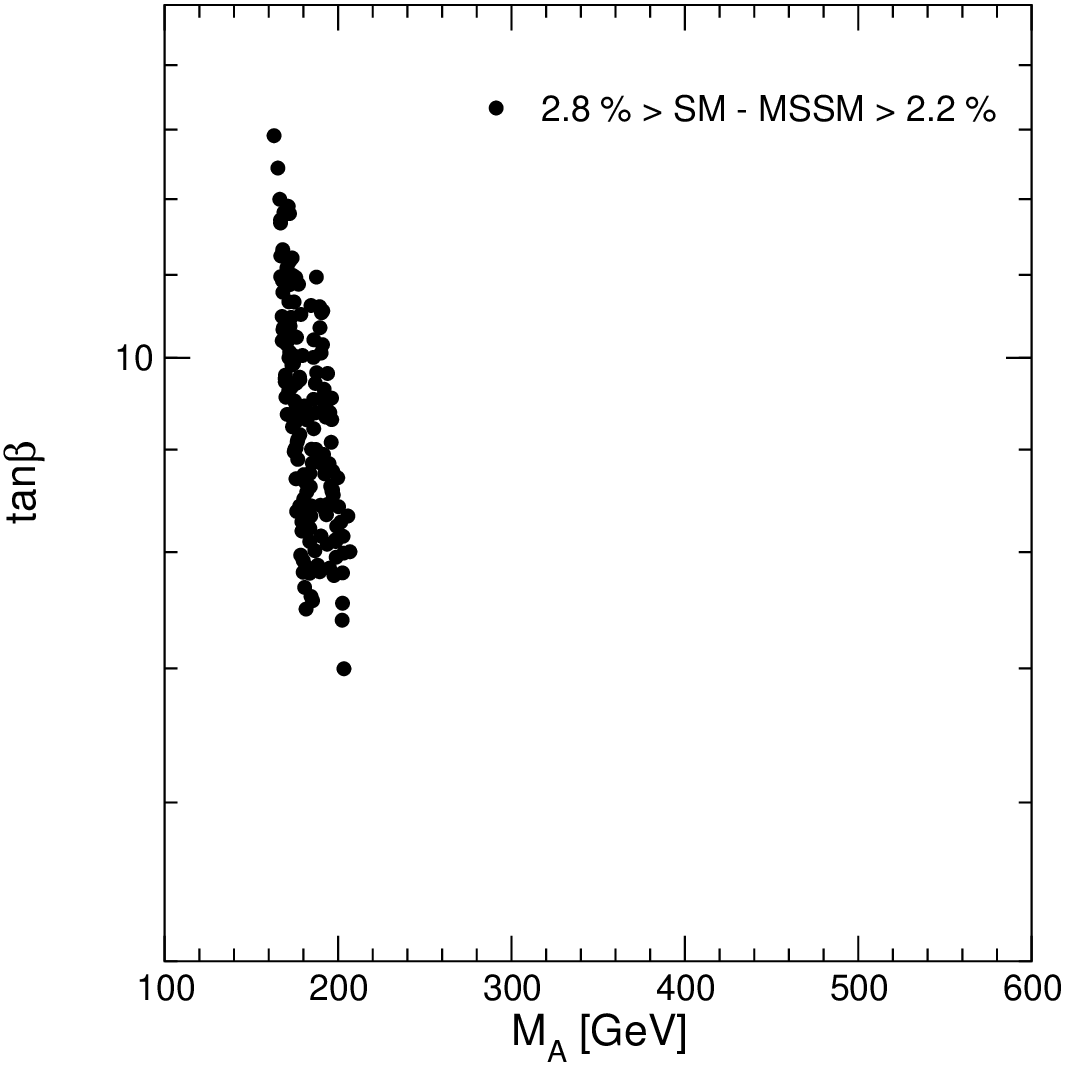,width=\linewidth,height=0.95\linewidth}}&
\mbox{\epsfig{file=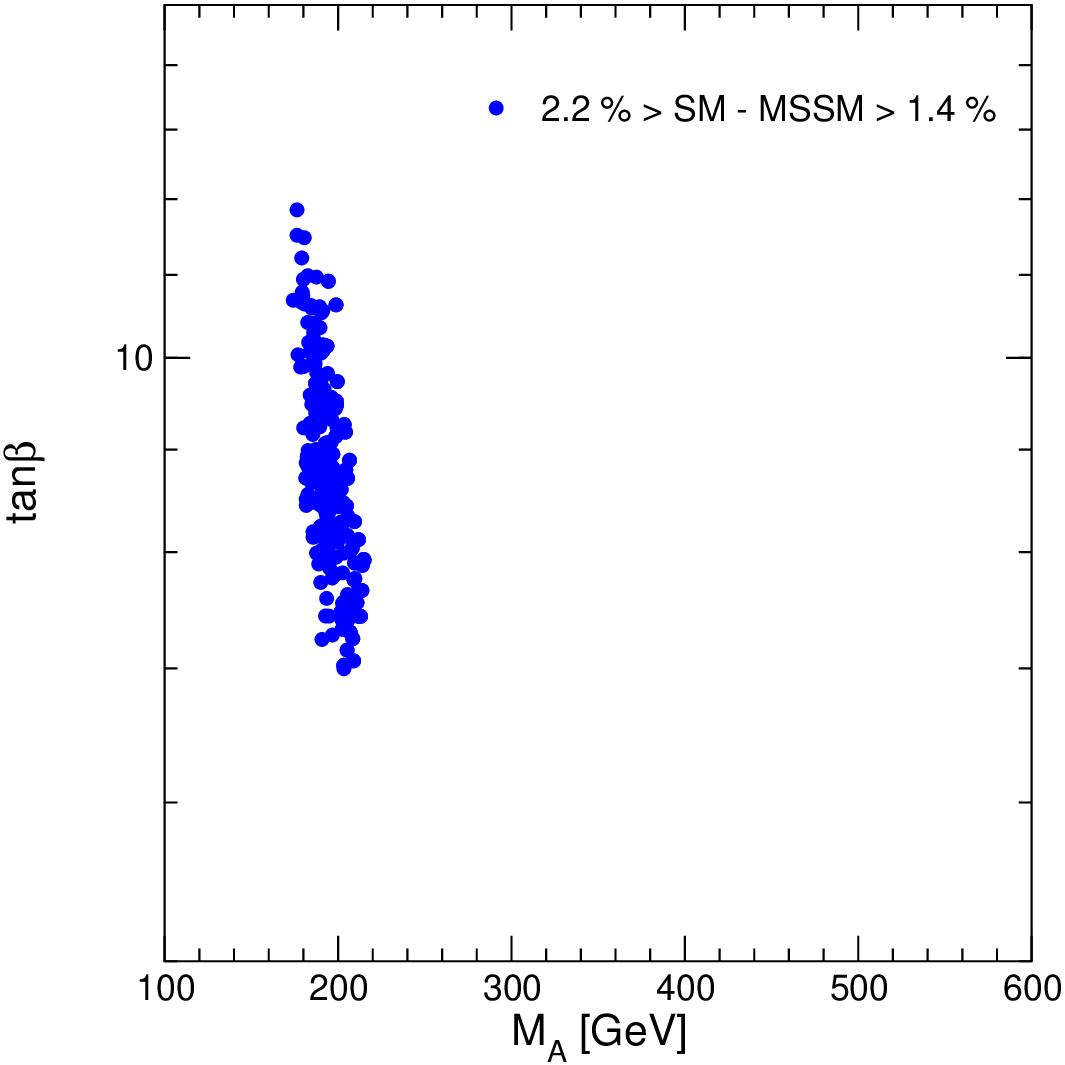,width=\linewidth,height=0.95\linewidth}}\\
\mbox{\epsfig{file=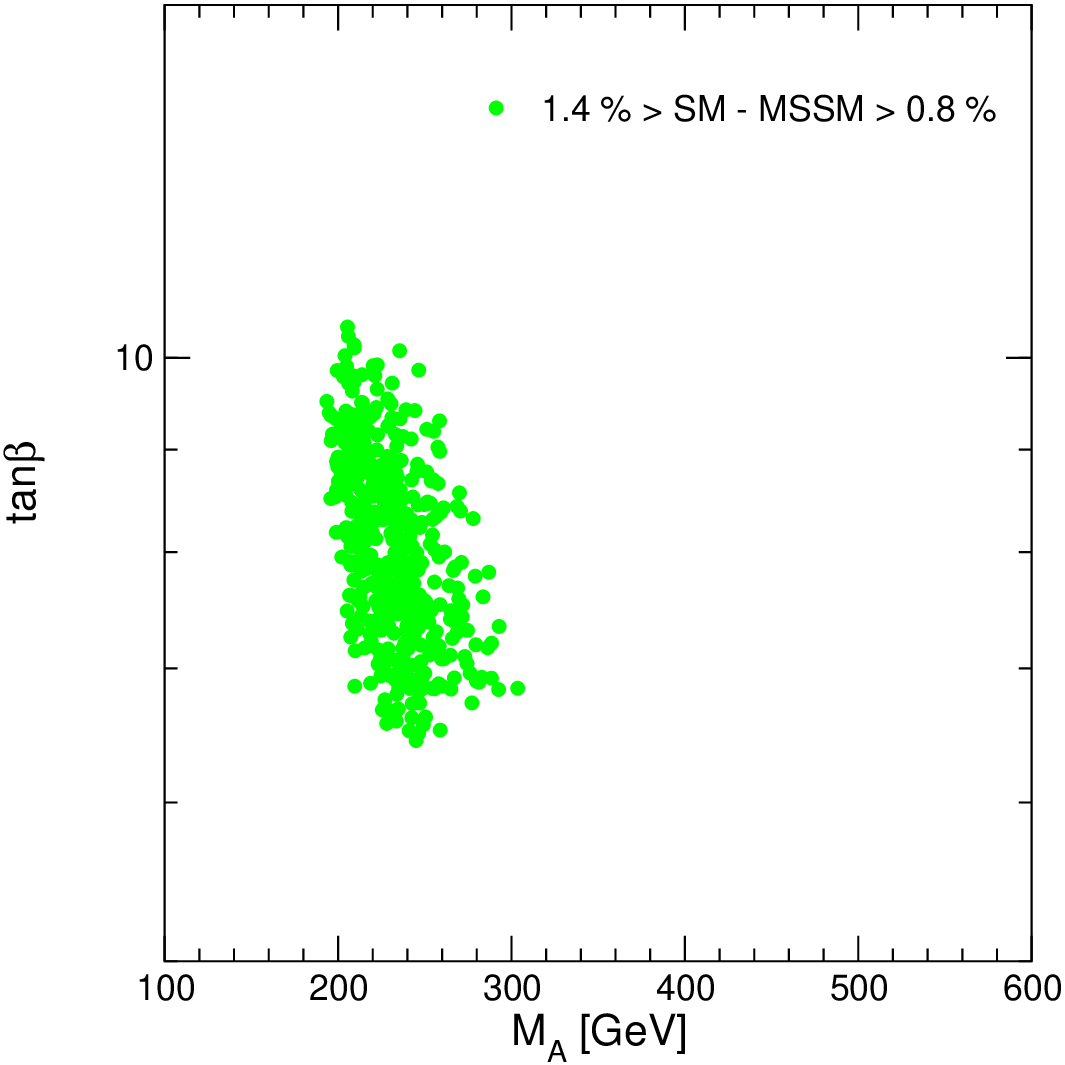,width=\linewidth,height=0.95\linewidth}}&
\mbox{\epsfig{file=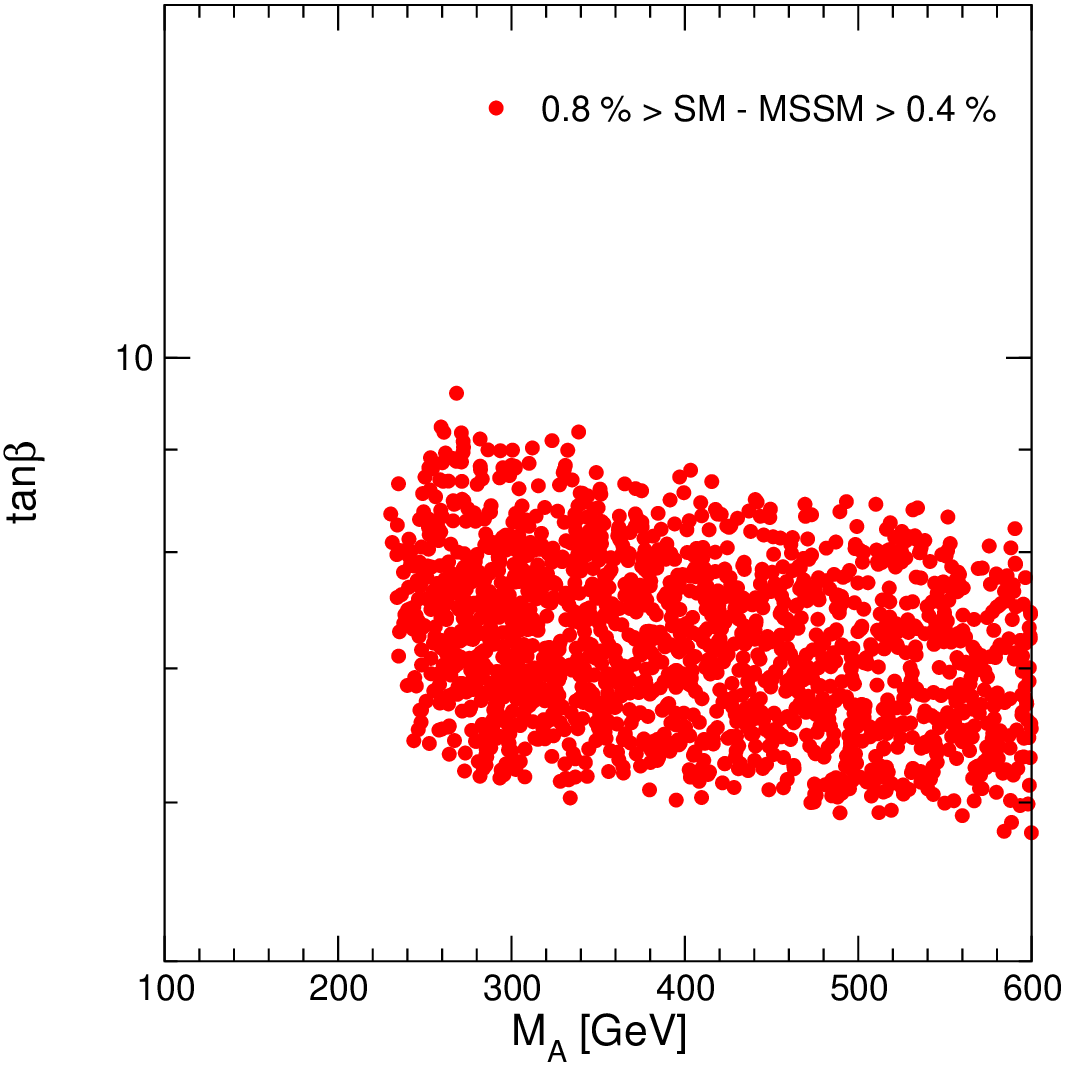,width=\linewidth,height=0.95\linewidth}}\\
\end{tabular}
\caption{The deviation of $\sihZ^{\SU}$ from 
$\sihZ^{\SM} = 0.1530~ {\rm pb}$ is shown in the $\MA-\tb$-plane
for $\Mh = 115 \pm 0.5 \gev$ at $\sqrt{s}=350~\gev$ with $\cL = 500$~\ifb.
For the $\Stop$~sector we have assumed the LC errors in \refeq{stopunc}.
}
\label{fig:ma_dep}
\end{center}
\end{figure}
%%%%%%%%%%%% F I G U R E %%%%%%%%%%%%%%%%%%%%%%%%%%%%%%%%%%%%%

\smallskip
With the above choices, the only remaining free parameters are $\MA$
and $\tb$, which we assume to be only poorly known in this scenario.
Our  procedure is to pick a value for $\MA$ and $\tb$ and check that
the chosen parameters generate $\Mh = 115 \pm 0.5 \gev$, which cuts
out a slice of the $\MA-\tb$-plane.
For the above set of parameters, we then calculate $\sihZ$ and compare 
with the value obtained for the SM, 
$\sihZ^{\SM} = 0.1530~{\rm pb}$.
The resulting variations of the cross section from the SM value are 
shown in \reffi{fig:ma_dep}.
Since the measurement of $\sihZ$ is a missing mass experiment, our
results are independent of the Higgs boson decay channel.

The different panels of \reffi{fig:ma_dep} show the regions where
the rate differs from the SM prediction by a specified amount.  
This includes a theoretical uncertainty
 in the SM rate which we approximate by varying
$\Mh$ within the range, 
$\Mh = 115 \pm 0.5 \gev$ (as described in \refse{sec:concept}).
The cross section is quite sensitive to $\tb$. 
A measurement which differs from the SM prediction by $1.4\%$ 
or less will restrict $\tb < 10$. 
Concerning the indirect $\MA$ determination, a $1.4\%$~measurement
would only be sensitive to $\MA \lsim 200 \gev$. However, a
measurement at the $0.5\%$~level, finding a deviation larger than
$0.8\%$, can be realized only for $\MA \lsim 300 \gev$.
A smaller deviation from the SM value can be realized for all $\MA$
values with $\MA \gsim 300 \gev$. Thus a weak upper
bound might be established; in case of a direct observation (which
will  be possible for such small $\MA$ values), the cross
section measurement can confirm the direct $\MA$ measurement.
Interestingly, this could also happen for values
where the LHC can see only the lightest MSSM Higgs boson (in the
so-called ``LHC wedge region''). 
The currently envisaged accuracy on $\sihZ$ of $2-3\%$ is 
unfortunately not sufficient for $\sihZ$ to be used as a such a
precision determination.

In this scenario it is important to keep the uncertainties of the
$\Stop$~sector in mind, which up to now we have assumed to come
partially from the LC and partially from the LHC, see
\refeq{stopunc}. If the more conservative assumption of LHC errors is
made, the cut-out region in the $\MA-\tb$-plane is visibly
enlarged. In particular the band is widened to larger $\tb$~values by
about~2, depending somewhat on $\MA$. The obtained results from the
cross section measurement for $\MA$ are affected in a two-fold way.
The lower $\MA$ bound is hardly affected at all. The upper $\MA$ bound
is weakened by $\sim 50 \gev$ in the relevant $\MA$ region, 
$\MA \sim 300 \gev$, but without spoiling the possible determination
of an upper bound as explained in the previous section.

%%%%%%%%%%%%%%%%%%%%%%%%%%%%%%%%%%%%%%%%%%%%%%%%%%%%%%%%%%%%%%
%%%%%%%%%%%%%%%%%%%%%%%%%%%%%%%%%%%%%%%%%%%%%%%%%%%%%%%%%%%%%%

\subsection{The gaugino scenario}
\label{subsec:gauginoscenario}

To demonstrate the possible amount of information that $\sihZ$ might
deliver on $\mu$
and $M_2$, in this scenario we make the assumption that $\MA$ and $\tb$
will have been measured, 

\BEA
\MA &=& 250 \pm 10 \gev \cr
\tb &=& 4 \pm 0.5 ~,
\EEA
but we leave the gaugino mass parameters
$M_2$ and $\mu$ as free parameters (the scan stops at an upper bound
of $1 \tev$). 
The other MSSM parameters are assumed to have the same values as in
\refse{subsec:higgsscenario}, together with their corresponding
uncertainties. As in the previous section, all experimental and
theoretical errors are fully taken into account.

%%%%%%%%%%%% F I G U R E %%%%%%%%%%%%%%%%%%%%%%%%%%%%%%%%%%%%%
\begin{figure}[htb!]
\begin{center}
\begin{tabular}{p{0.48\linewidth}p{0.48\linewidth}}
\mbox{\epsfig{file=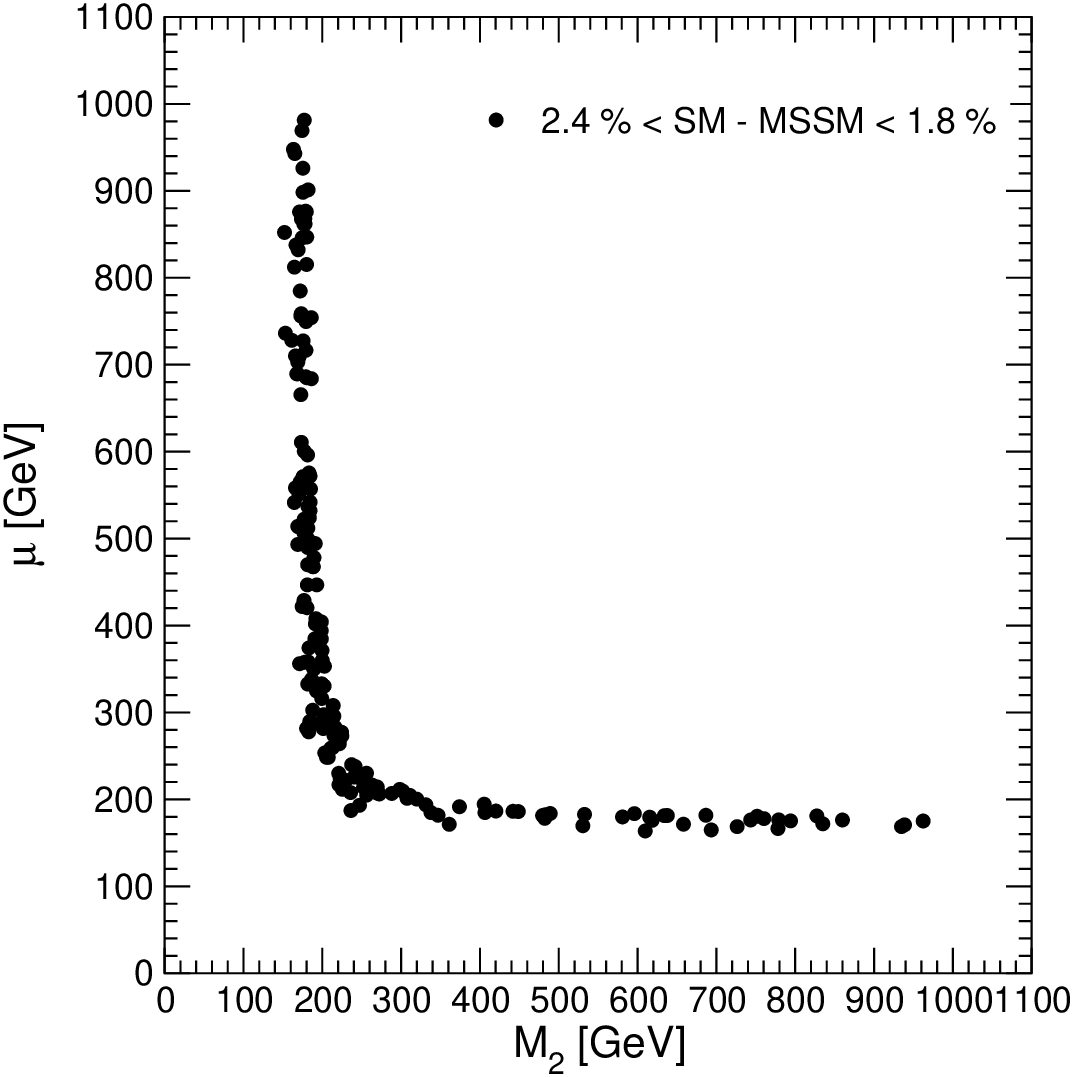,width=\linewidth,height=0.95\linewidth}}&
\mbox{\epsfig{file=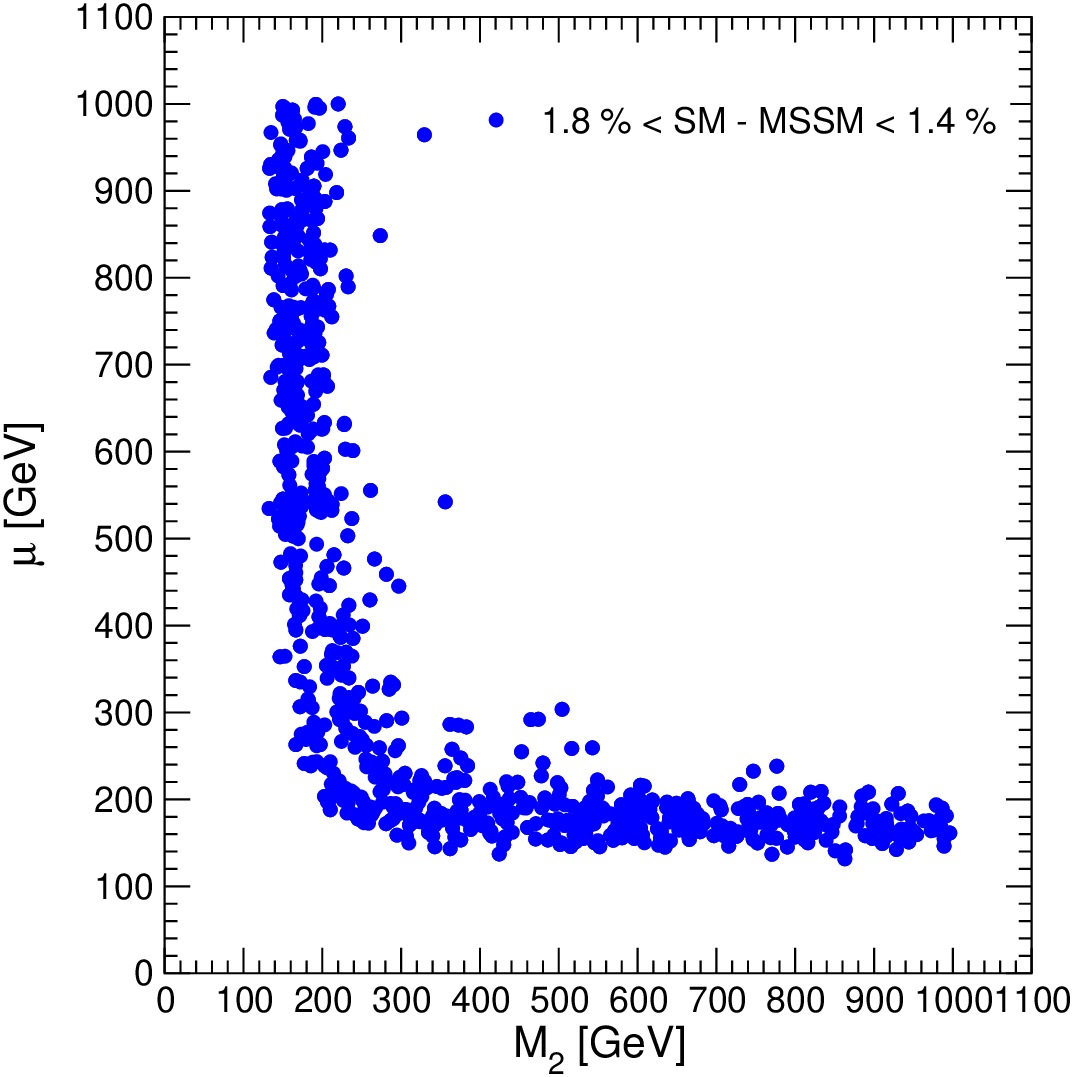,width=\linewidth,height=0.95\linewidth}}\\
\mbox{\epsfig{file=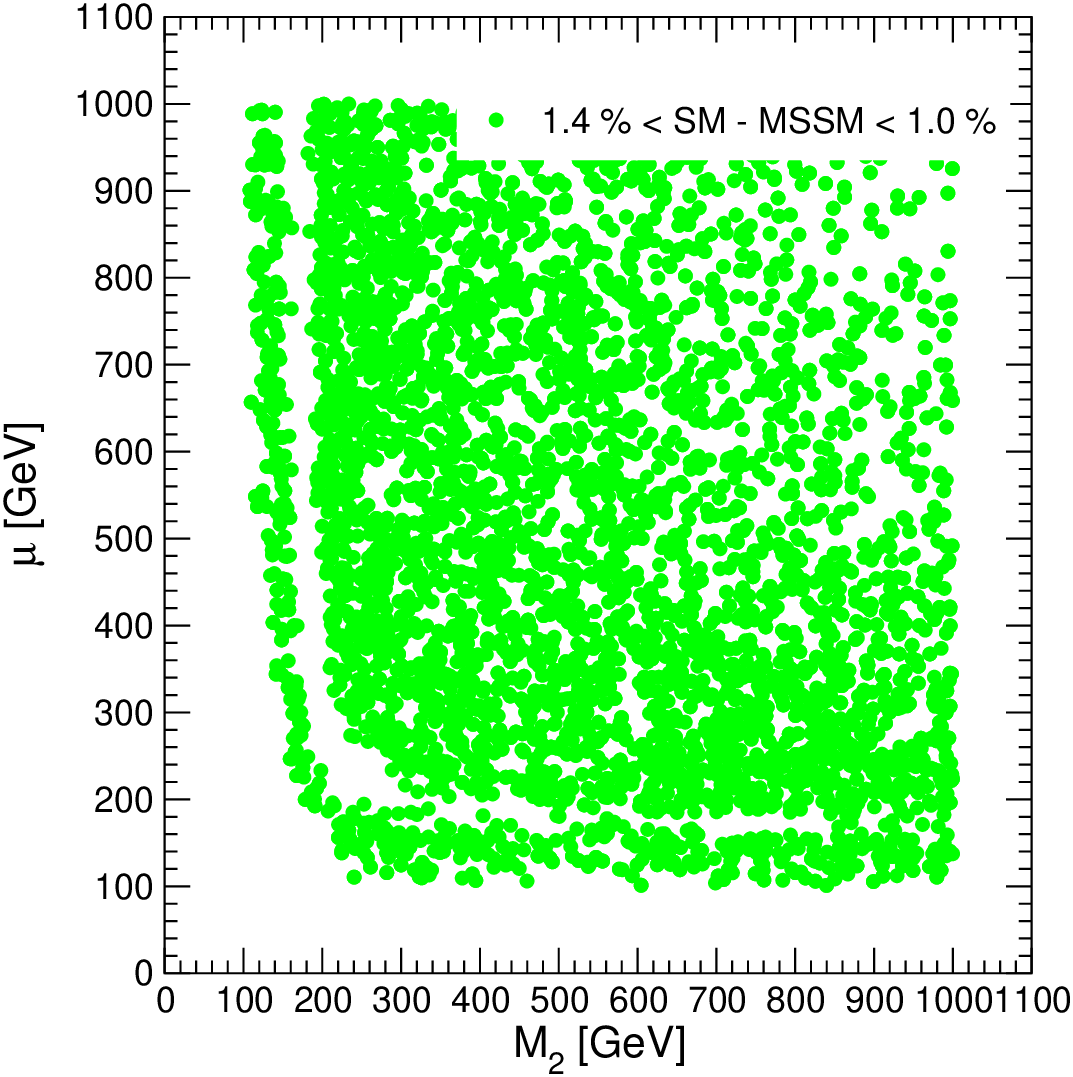,width=\linewidth,height=0.95\linewidth}}&
\mbox{\epsfig{file=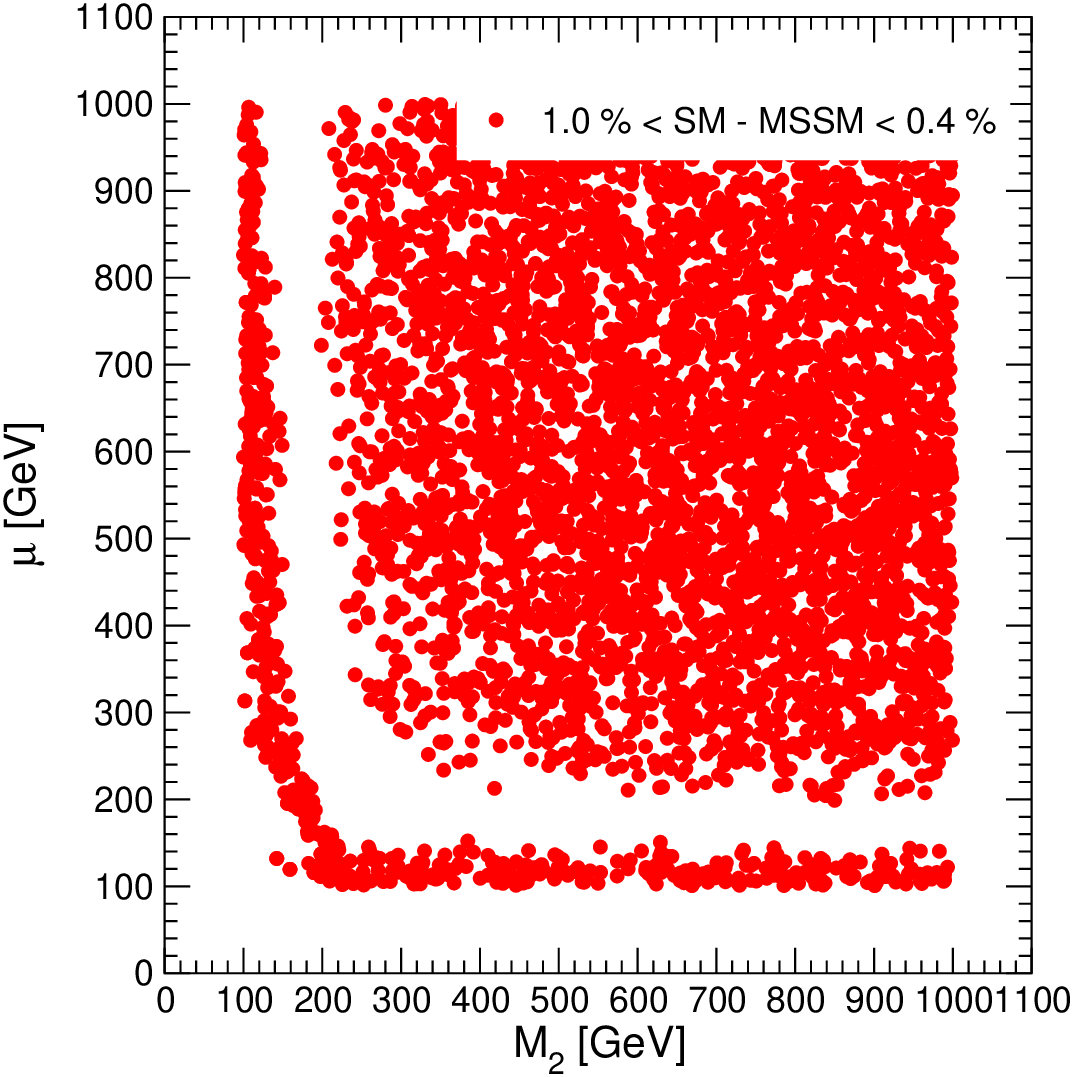,width=\linewidth,height=0.95\linewidth}}\\
\end{tabular}
\caption{The deviation of $\sihZ^{\SU}$ from 
$\sihZ^{\SM} = 0.1530~ {\rm pb}$ is shown in the $M_2-\mu$-plane for
$\Mh = 115 \pm 0.5 \gev$ at
$\sqrt{s}=350~\gev$ with $\cL = 500$~\ifb.
}
\label{fig:m2dep}
\end{center}
\end{figure}
%%%%%%%%%%%% F I G U R E %%%%%%%%%%%%%%%%%%%%%%%%%%%%%%%%%%%%%

\reffi{fig:m2dep} shows the dependence of $\sihZ^{\SU}$ on $\mu$ and
$M_2$. It is obvious that a reasonable sensitivity only appears for
$M_2 \approx 200 \gev$ or $\mu \approx 200 \gev$, where
$\sihZ^{\SU}$ has a minimum. It is very unlikely that these two
parameters, if they possess such a low value, will not have been
measured directly, see e.g.~\cite{teslatdr} and references
therein. Thus, in this scenario $\sihZ$ can only offer 
complementary information which  can verify the internal consistency of
the MSSM (see \citere{gigaz} for detailed discussion on this
subject).

%%%%%%%%%%%%%%%%%%%%%%%%%%%%%%%%%%%%%%%%%%%%%%%%%%%%%%%%%%%%%%
%%%%%%%%%%%%%%%%%%%%%%%%%%%%%%%%%%%%%%%%%%%%%%%%%%%%%%%%%%%%%%

\section{Conclusions and Outlook}

We have investigated whether a precise measurement of the Higgs production
cross section, $\si(e^+e^- \to hZ)$, offers additional information to
pin down the unknown parameters of the MSSM. We have chosen two
possible future scenarios. We have explained in detail what
uncertainties will be present at the time of a $\sihZ$ measurement and
how we take them into account. This includes
realistic assumptions for all 
mass parameters together with the expected uncertainties obtainable at
the LHC and/or LC. We also took into account realistic assumptions on
the theoretical uncertainties for the predictions in the MSSM Higgs
sector. 

We find that the total rate needs to be 
measured to a $0.5 - 1\%$ accuracy in order to
be useful as a precision observable. Then additional information on
$\tb$ or $\MA$ (if it is not too high, $\MA \lsim 500 \gev$) may be
obtainable. The dependence of $\sihZ$ on the gaugino parameters $\mu$
and $M_2$ shows a strong enough dependence
to be useful only for very low values,
$\mu, M_2 \approx 200 \gev$. Hence, in this case $\sihZ$ could only
test the internal consistency of the MSSM.

The required precision for $\sihZ$ at the 1\%~level, as compared to
the envisaged $2 - 3\%$, could possibly achieved by either
accumulating a higher integrated luminosity (also at different center
of mass energies) and/or by taking other than the leptonic $Z$~decay modes
into account. 

%%%%%%%%%%%%%%%%%%%%%%%%%%%%%%%%%%%%%%%%%%%%%%%%%%%%%%%%%%%%%%
%%%%%%%%%%%%%%%%%%%%%%%%%%%%%%%%%%%%%%%%%%%%%%%%%%%%%%%%%%%%%%

\subsection*{Acknowledgement}
This work supported by the Department of Energy under contract 
DE-AC02-98CH10886.

%%%%%%%%%%%%%%%%%%%%%%%%%%%%%%%%%%%%%%%%%%%%%%%%%%%%%%%%%%%%%%
%%%%%%%%%%%%%%%%%%%%%%%%%%%%%%%%%%%%%%%%%%%%%%%%%%%%%%%%%%%%%%


\begin{thebibliography}{99}

\bibitem{higgstev} M.~Carena, J.~Conway, H.~Haber and J.~Hobbs et al., 
                   [Tevatron Higgs working group],
                   hep-ph/0010338.
                   %%CITATION = HEP-PH 0010338;%%

\bibitem{atlastdr} ATLAS Collaboration, 
  {\em Detector and Physics Performance Technical Design Report},
  CERN/LHCC/99-15 (1999), see:\\
 {\tt atlasinfo.cern.ch/Atlas/GROUPS/PHYSICS/TDR/access.html} .

\bibitem{cms} CMS Collaboration, see:\\
 {\tt cmsinfo.cern.ch/Welcome.html/CMSdocuments/CMSplots/} .

\bibitem{ehow} J.~Ellis, S.~Heinemeyer, K.~Olive and G.~Weiglein,
               {\em Phys. Lett.} {\bf  B515 } (2001) 348,
               hep-ph/0105061.
               %%CITATION = HEP-PH 0105061;%%

\bibitem{hhh} A.~Djouadi, W.~Kilian, M.~M\"uhlleitner and P.M.~Zerwas,
              {\em Eur. Phys. Jour.} {\bf C 10} (1999) 45, 
              hep-ph/9903229;\\
              %%CITATION = HEP-PH 09903229;%%
              F.~Boudjema and A.~Semenov,
              hep-ph/0201219.
              %%CITATION = HEP-PH 0201219;%%

\bibitem{teslatdr} J.~Aguilar-Saavedra et al. 
                   [ECFA/DESY LC Physics Working Group Collaboration],
                   ``TESLA Technical Design Report Part III: Physics
                   at an e+e- Linear Collider'',
                   hep-ph/0106315,
                   %%CITATION = HEP-PH 0106315;%%
                   see: {\tt tesla.desy.de} .

\bibitem{orangebook} T.~Abe {\it et al.}  
                     [American LC Working Group Collaboration],
                     {\it Resource book for Snowmass 2001}, 
                     hep-ex/0106056.
                     %%CITATION = HEP-EX 0106056;%%

\bibitem{acfarep} K.~Abe et al. 
                  [ACFA LC Working Group Collaboration],
                  hep-ph/0109166.
                  %%CITATION = HEP-PH 0109166;%%

\bibitem{lepewwg} LEP Electroweak Working group,\\ 
                  see {\tt http://lepewwg.web.cern.ch/LEPEWWG/Welcome.html}

\bibitem{blueband} U.~Baur, R.~Clare, J.~Erler, S.~Heinemeyer,
                   D.~Wackeroth, G.~Weiglein and D.~Wood,
                   contribution to the P1-WG1 report of the workshop
                   ``The Future of Particle Physics'', Snowmass, 
                   Colorado, USA, July 2001,
                   hep-ph/0111314.
                   %%CITATION = HEP-PH 0111314;%%

\bibitem{hprod} W.~Kilian, M.~Kr\"amer and P.M.~Zerwas,
                {\em Phys. Lett.} {\bf B 373} (1996) 135,
                hep-ph/9512335.
                %%CITATION = HEP-PH 9512335;%%

\bibitem{brau} J.~Brau, C.~Potter and M.~Iwasaki,
               published in ``Batavia 2000, Physics and experiments
               with future linear e+ e- colliders'', p267.

\bibitem{zeppi} D.~Zeppenfeld, R.~Kinnunen, A.~Nikitenko and 
                E.~Richter-Was, 
                {\em Phys. Rev.} {\bf D 62} (2000) 013009,
                hep-ph/0002036.
                %%CITATION = HEP-PH 0002036;%%

\bibitem{mssm} H.P.~Nilles, 
               {\em Phys.\ Rep.} {\bf 110} (1984) 1; \\ 
               %%CITATION = PRPLC,110,1;%%
               H.E.~Haber and G.L.~Kane, 
               {\em Phys.\ Rep.} {\bf 117}, (1985) 75; \\  
               %%CITATION = PRPLC,117,75;%%
               R.~Barbieri, 
               {\em Riv.\ Nuovo Cim.} {\bf 11}, (1988) 1. 
               %%CITATION = RNCIB,11,1;%%

\bibitem{higgsbrimp} K.~Babu and C.~Kolda,
                     {\em Phys. Lett.} {\bf B 451} (1999) 77,
                     hep-ph/9811308;\\
                     %%CITATION = HEP-PH 9811308;%%
                     S.~Heinemeyer, W.~Hollik and G.~Weiglein
                     {\em Eur. Phys. Jour.} {\bf C 16} (2000) 139,
                     hep-ph/0003022;\\
                     %%CITATION = HEP-PH 0003022;%%
                     S.~Heinemeyer and G.~Weiglein, 
                     hep-ph/0102117.
                     %%CITATION = HEP-PH 0102117;%%

\bibitem{higgsbrimp2} J.~Guasch, W.~Hollik and S.~Pe\~naranda,
                      {\em Phys. Lett.} {\bf B 515} (2001) 367,
                      hep-ph/0106027;\\
                      %%CITATION = HEP-PH 0106027;%%
                      M.~Carena, H.~Haber, H.~Logan and S.~Mrenna,
                      {\em Phys. Rev.} {\bf D 65} (2002) 055005, 
                      hep-ph/0106116;\\
                      %%CITATION = HEP-PH 0106116;%%
                      A.~Curiel, M.~Herrero, D.~Temes and J.~De Troconiz,
                      {\em Phys. Rev.} {\bf D 65} (2002) 075006,
                      hep-ph/0106267.
                      %%CITATION = HEP-PH 0106267;%%

\bibitem{eehZ1l} P.H. Chankowski, S. Pokorski and J. Rosiek, 
                 {\em Nucl. Phys.} {\bf B 423} (1994) 437,
                 hep-ph/9303309;
                 %%CITATION = HEP-PH 9303309;%%
                 {\em Nucl. Phys.} {\bf B 423} (1994) 497;\\
                 %%CITATION = NUPHA,B423,497;%%
                 V.~Driesen and W.~Hollik,
                 {\em Z. Phys.} {\bf C 68} (1995), 485,
                 hep-ph/9504335;\\
                 %%CITATION = HEP-PH 9504335;%%
                 V. Driesen, W. Hollik and J.  Rosiek, 
                 {\em Z. Phys.} {\bf C 71} (1996) 259,
                 hep-ph/9512441.
                 %%CITATION = HEP-PH 9512441;%%

\bibitem{eehZ2l} S.~Heinemeyer, W.~Hollik, J.~Rosiek and G.~Weiglein, 
                {\em Eur. Phys. J.} {\bf C 19} (2001) 535,
                hep-ph/0102081;\\
                %%CITATION = HEP-PH 0102081;%%
                S.~Heinemeyer and G.~Weiglein,
                {\em Nucl. Phys. Proc. Suppl.} {\bf 89} (2000) 210.
                %%CITATION = NUPHZ,89,210;%%

\bibitem{hhg} J.~Gunion, H.~Haber, G.~Kane and S.~Dawson,
              {\em The Higgs Hunter's Guide}, Addison-Wesley, 1990.

\bibitem{mhiggs2lA} M.~Carena, J.~Espinosa, M.~Quir\'os and C.~Wagner, 
                    {\em Phys. Lett.} {\bf B 355} (1995) 209, 
                    hep-ph/9504316;\\
                    %%CITATION = HEP-PH 9504316;%%
                    M.~Carena, M.~Quir\'os and C.~Wagner, 
                    {\em Nucl. Phys.} {\bf B 461} (1996) 407, 
                    hep-ph/9508343;\\
                    %%CITATION = HEP-PH 9508343;%%
                    H.~Haber, R.~Hempfling and A.~Hoang, 
                    {\em Z. Phys.} {\bf C 75} (1997) 539, 
                    hep-ph/9609331.
                    %%CITATION = HEP-PH 9609331;%%
                    R.~Zhang, 
                    {\em Phys. Lett.} {\bf B 447} (1999) 89, 
                    hep-ph/9808299;\\
                    %%CITATION = HEP-PH 9808299;%%
                    J.~Espinosa and R.~Zhang,
                    {\em JHEP} {\bf 0003} (2000) 026, 
                    hep-ph/9912236;
                    %%CITATION = HEP-PH 9912236;%%
                    {\em Nucl. Phys.} {\bf B 586} (2000) 3,
                    hep-ph/0003246;\\
                    %%CITATION = HEP-PH 0003246;%%
                    G.~Degrassi, P.~Slavich and F.~Zwirner,
                    {\em Nucl. Phys.} {\bf B 611} (2001) 403,
                    hep-ph/0105096;\\
                    %%CITATION = HEP-PH 0105096;%%
                    A.~Brignole, G.~Degrassi, P.~Slavich and F.~Zwirner,
                    {\em Nucl. Phys.} {\bf B 631} (2002) 195, 
                    hep-ph/0112177;
                    %%CITATION = HEP-PH 0112177;%%
                    hep-ph/0206101.
                    %%CITATION = HEP-PH 0206101;%%

\bibitem{mhiggs2lB} S.~Heinemeyer, W.~Hollik and G.~Weiglein, 
                    {\em Phys. Rev.} {\bf D 58} (1998) 091701, 
                    hep-ph/9803277; 
                    %%CITATION = HEP-PH 9803277;%%
                    {\em Phys. Lett.} {\bf B 440} (1998) 296, 
                    hep-ph/9807423.
                    %%CITATION = HEP-PH 9807423;%%
                    {\em Eur. Phys. Jour.} {\bf C 9} (1999) 343, 
                    hep-ph/9812472.
                    %%CITATION = HEP-PH 9812472;%%

\bibitem{bse} M.~Carena, H.~Haber, S.~Heinemeyer, W.~Hollik, C.~Wagner
              and G.~Weiglein, 
              {\em Nucl. Phys.} {\bf B 580} (2000) 29, 
              hep-ph/0001002.
              %%CITATION = HEP-PH 0001002;%%

\bibitem{mhiggs1l} H.~Haber and R.~Hempfling,
                   {\em Phys. Rev. Lett.} {\bf 66} (1991) 1815;\\
                   %%CITATION = PRLTA,66,1815;%%
                   Y.~Okada, M.~Yamaguchi and T.~Yanagida,
                   {\em Prog. Theor. Phys.} {\bf 85} (1991) 1;\\
                   %%CITATION = PTPKA,85,1;%%
                   J.~Ellis, G.~Ridolfi and F.~Zwirner,
                   {\em Phys. Lett.} {\bf B 257} (1991) 83; 
                   %%CITATION = PHLTA,B257,83;%%
                   {\em Phys. Lett.} {\bf B 262} (1991) 477;\\
                   %%CITATION = PHLTA,B262,477;%%
                   R.~Barbieri and M.~Frigeni,
                   {\em Phys. Lett.} {\bf B 258} (1991) 395.
                   %%CITATION = PHLTA,B258,395;%%

\bibitem{gigaz} J.~Erler, S.~Heinemeyer, W.~Hollik, G.~Weiglein 
                and P.M.~Zerwas,
                {\em Phys. Lett.} {\bf B 486} (2000) 125,
                hep-ph/0005024.
                %%CITATION = HEP-PH 0005024;%%

\bibitem{runscen} J.~Barron et al.,
                  contribution to the workshop
                  ``The Future of Particle Physics'', Snowmass, 
                  Colorado, USA, July 2001,
                  hep-ph/0201177.
                  %%CITATION = HEP-PH 0201177;%%

\bibitem{dmhtheo} M.~Frank, S.~Heinemeyer, W.~Hollik and G.~Weiglein,
                  hep-ph/0202166.
                  %%CITATION = HEP-PH 0202166;%%

\bibitem{mtdet} A.~Hoang et al., 
                {\em Eur. Phys. J.} {\bf C 3} (2000) 1,
                hep-ph/0001286;\\
                %%CITATION = HEP-PH 0001286;%%
                M.~Martinez,
                talk at the Linear Collider Workshop, Cracow, Poland,
                September 2001,\\
                see: {\tt webnt.physics.ox.ac.uk/lc/ecfadesy\_topqcd.htm}~;\\
                M.~Martinez and R.~Miquel,
                hep-ph/0207315.
                %%CITATION = HEP-PH 0207315;%%

\bibitem{feynhiggs} S.~Heinemeyer, W.~Hollik and G.~Weiglein, 
                    {\em Comp. Phys. Comm.} {\bf 124} (2000) 76,
                    hep-ph/9812320; 
                    %%CITATION = HEP-PH 9812320;%%%
                    hep-ph/0002213.
                    %%CITATION = HEP-PH 0002213;%%
                    The codes are accessible at 
                    {\tt www.feynhiggs.de} .

\bibitem{LHCstop} M.~Nojiri, 
                  talk at the ``SUSY02'', DESY, Hamburg, Germany,
                  June 2002,\\ see:
                  {\tt www.desy.de/susy02/susy02\_parallel\_1B.html}

\end{thebibliography}
\end{document}